\pdfoutput=1
\documentclass[10pt,conference]{IEEEtran}
\IEEEoverridecommandlockouts\IEEEpubid{\makebox[\columnwidth]{ 978-1-6654-3540-6/22~\copyright~2022 IEEE \hfill} \hspace{\columnsep}\makebox[\columnwidth]{ }}
\usepackage[numbers,sort&compress]{natbib}

\usepackage{comment}

\makeatletter
\usepackage{mathtools}
\usepackage{amsmath}

\usepackage{makecell}
\usepackage{float}
\usepackage{graphicx,times,amsmath,booktabs,algorithm,algorithmic,amsmath}
\usepackage{multicol}
\usepackage{amsmath,amssymb}
\usepackage{subfigure}
\usepackage{stfloats}

\begin{document}
	\title{Spectral Efficiency Analysis of Uplink-Downlink Decoupled Access in C-V2X Networks}

	\author{\IEEEauthorblockN{Luofang Jiao\IEEEauthorrefmark{1}, Kai Yu\IEEEauthorrefmark{1}, Yunting~Xu\IEEEauthorrefmark{1}, Tianqi~Zhang\IEEEauthorrefmark{1}, Haibo Zhou\IEEEauthorrefmark{1}, and Xuemin (Sherman) Shen\IEEEauthorrefmark{2}}
	\IEEEauthorblockA{\IEEEauthorrefmark{1}School of Electronic Science and Engineering, Nanjing University, Nanjing, China, 210023}
	Email: {luofang\_jiao@foxmail.com, \{kaiyu, yuntingxu, tianqizhang\}@smail.nju.edu.cn, haibozhou@nju.edu.cn}\\
    \IEEEauthorblockA{\IEEEauthorrefmark{2}Department of Electrical and Computer Engineering, University of Waterloo, 200 University Avenue West, \\Waterloo, Ontario, Canada, N2L 3G1. Email: {sshen@uwaterloo.ca}}
}

\maketitle
\begin{abstract}

The uplink (UL)/downlink (DL) decoupled access has been emerging as a novel access
architecture to improve the performance gains in cellular networks. In this paper, we investigate the
UL/DL decoupled access performance in cellular vehicle-to-everything (C-V2X).
We propose a unified analytical framework for the UL/DL decoupled access in C-V2X from the perspective
of spectral efficiency (SE). By modeling the UL/DL decoupled access C-V2X as a Cox process and leveraging the stochastic geometry, we obtain the joint association probability, the UL/DL distance distributions to serving base stations and the SE for the UL/DL decoupled access in C-V2X
networks with different association cases. We conduct extensive Monte Carlo
simulations to verify the accuracy of the proposed
unified analytical framework, and the results show a better system average SE
of UL/DL decoupled access in C-V2X.

\end{abstract}
\IEEEpeerreviewmaketitle

\begin{IEEEkeywords}
C-V2X, uplink/downlink decoupled access, association probability, spectral efficiency, stochastic geometry.
\end{IEEEkeywords}

\section{Introduction}

\IEEEPARstart W{ith} the rapid development of the cellular vehicle-to-everything (C-V2X) networks,
it is of critical significance to enable high quality of service (QoS) of uplink (UL) and downlink (DL) for supporting the emerging advanced vehicular
applications, such as stream media, autonomous vehicles and
intelligent transportation systems (ITS) \cite{gu2016exploiting, abboud2016interworking}.
Nevertheless, to satisfy the ever-increasing coverage demand and the more stringent service requirements of vehicle users, C-V2X is gradually evolving into a more complicated heterogeneous network composition, which is consisted of the macro base stations (MBS) and the small base stations (SBS) \cite{boccardi2016decouple, zhou2020evolutionary}.
However, the traditional user association and access mode suffer frequent network handover and low throughput for C-V2X networks \cite{zhou2020evolutionary}. It is widely accepted that the users located on the edge of SBS choose to access SBS in UL and access to MBS in DL can significantly increase UL rates in heterogeneous networks in cellular networks \cite{li2018uplink, yu2020reinforcement}. It is critical for the high UL rates guarantee such as sharing of security information and remote driving in C-V2X \cite{rabieh2018srpv, yao2019bla}. Therefore, it is meaningful to study the impact of novel access technologies with flexible user association in heterogeneous C-V2X networks.

Different from the traditional coupled access technologies, the UL/DL decoupled access technology, where the UL and DL separately access to any two different tiers of base stations (BS) (i.e., MBS and SBS), has been proved that it can bring significant gains in terms of coverage, throughput and load balancing, etc. \cite{sattar2019spectral}.
E.g., Zhang \textit{et al}. theoretically validated the UL performance improvement brought about by the UL/DL decoupled access over the conventional coupled UL/DL access mode \cite{zhang2017uplink}. 
Sattar \textit{et al}. in \cite{sattar2019spectral} made an in-depth analysis of spectral efficiency
(SE) of all links from the perspective of joint association cases leveraging stochastic geometry.
Furthermore, regarding UL/DL decoupled access in C-V2X networks,  Yu \textit{et al}. introduced the UL/DL decoupled access into C-V2X for the first time and proved that this technology could improve the UL's throughput and load balancing in C-V2X \cite{ yu2020reinforcement}.

Although the UL/DL decoupled access has emerged as a novel and flexible access scheme to improve the C-V2X network performance, how to conduct a unified and joint UL/DL analysis is critical for evaluating the quality of emerging advanced C-V2X applications. Generally, the joint association probability, the UL/DL distance distributions to serving BSs and the SE of the UL/DL decoupled access in C-V2X networks are of important research significance and attract lots of academic attention. And in C-V2X, Chetlur \textit{et al}. modeled the C-V2X as a Cox process and characterized the
coverage probability and rate in \cite{chetlur2019coverage}. Sial \textit{et al}. presented a model of
V2X over shared channels based on stochastic geometry \cite{8859331}.

In this paper, we propose a unified analytical framework and investigate the characteristics of the UL/DL decoupled access by leveraging the tool of stochastic geometry in C-V2X networks. We first model the C-V2X networks as a Cox process. Subsequently, we leverage the stochastic geometry to derive the joint association probability and distance distributions to serving BSs. Moreover, the expressions of SE for the UL/DL decoupled access in C-V2X networks with different association cases are derived. We highlight our contributions in this paper as follows:
\begin{itemize}
	\item We introduce the UL/DL decoupled access technology into C-V2X networks to better support C-V2X applications. Through modeling the UL/DL decoupled access C-V2X as a Cox process, a unified analytical framework is presented.

	\item We provide an in-depth theoretical analysis by the stochastic geometry, including the joint association probability, the UL/DL distance distributions to serving BSs, and the mathematical expressions of SE for different association cases in C-V2X networks.

	\item We conduct extensive Monte Carlo experimentation to verify the effectiveness of the proposed unified analytical framework and the simulation results provide a general guidance for the UL/DL decoupled access scheme in C-V2X networks.
\end{itemize}

The rest of this paper is organized as follows. Section \uppercase\expandafter{\romannumeral2} introduces the network model. In Section \uppercase\expandafter{\romannumeral3}, we give the problem formulation, analysis, solutions and proofs. The numerical and simulation results are presented and discussed in Section \uppercase\expandafter{\romannumeral4}. Finally, we conclude this paper in Section \uppercase\expandafter{\romannumeral5}.

\vspace{-0.01cm}
\section{SYSTEM MODEL}

\subsection{Network Model}

   We model a two-tier UL/DL decoupled C-V2X network, which consists of SBS and MBS deployed in terms of an independent homogeneous Poisson point process (PPP) $\Phi$  of density $\lambda$ in the Euclidean plane.

\begin{figure}[t]
	\centering
	\centerline{\includegraphics[scale=0.31]{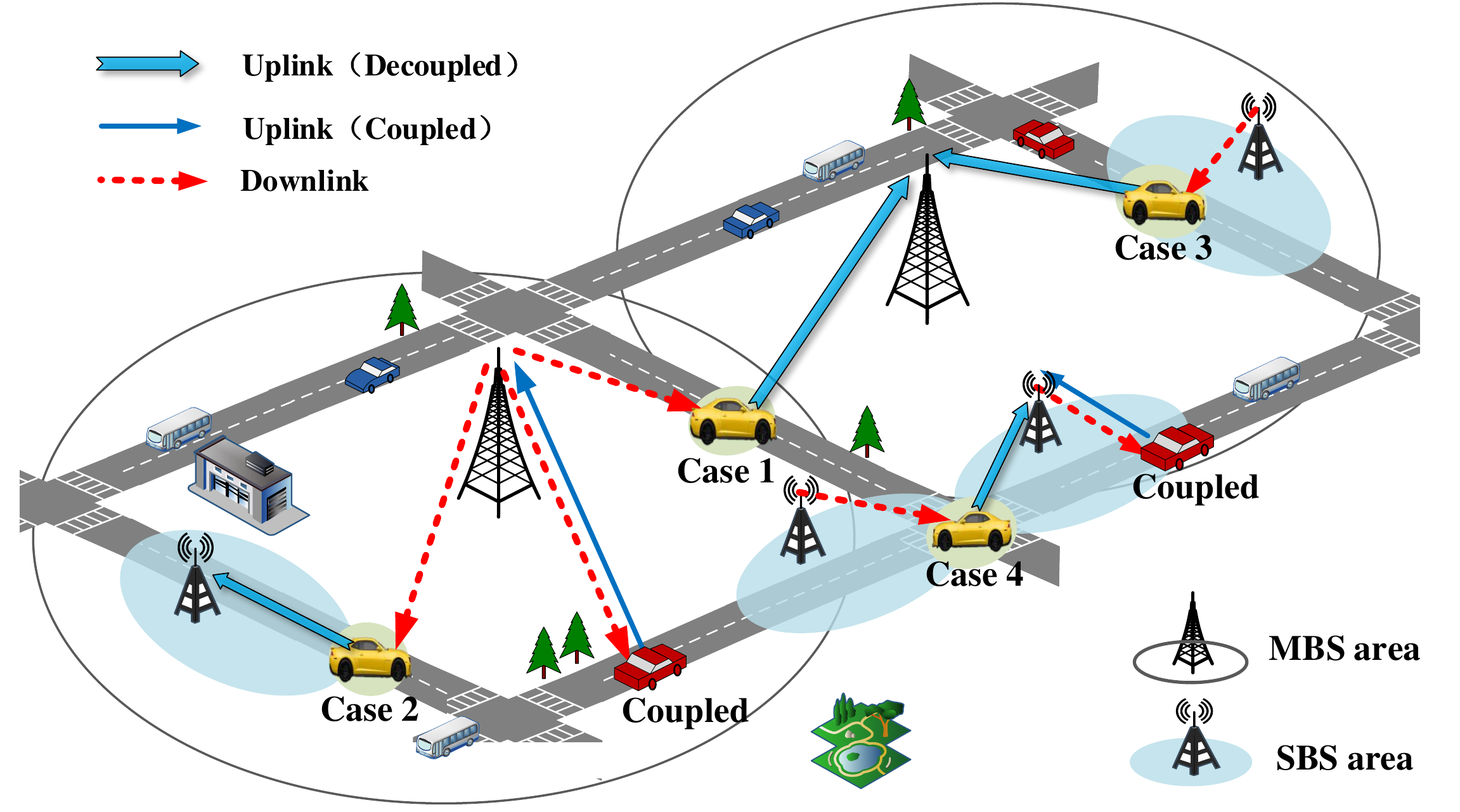}}
	\caption{Illustration of UL/DL decoupled access in C-V2X networks.}	
	\label{sys_model}
\end{figure}

   As shown in Fig.\ref{sys_model}, the UL/DL decoupled association process has four cases:
   \begin{itemize}
   	\item Case 1: UL = MBS, DL = MBS
   	\item Case 2: UL = SBS, DL = MBS
   	\item Case 3: UL = MBS, DL = SBS
   	\item Case 4: UL = SBS, DL = SBS
   \end{itemize}

According to the characteristics of C-V2X \cite{chetlur2019coverage}, the MBSs are evenly distributed in the C-V2X networks while the SBSs are distributed along the roads. All BSs are deployed in a circular area with a radius of $r$. A road that passes through the origin is called the typical line and the roads that don't pass through the origin are called the other lines. We choose a vehicle on the typical lines as the typical vehicle. ${\Phi _n}$ denotes the set of BSs in terms of PPP with density ${\lambda _n} $, where $n\in \left\{ {M,S} \right\}$ for MBS and SBS, respectively. We use $  \Xi _{{l_0}}^{(S)} $ to denote the set of vehicles in the typical line $ l_0 $. The locations of vehicles and BSs form a double stochastic Poisson process which is called a Cox processes.

The MBS chooses to transmit to vehicles in typical line and other lines in the same power. The SBS uses beamforming and the directional antenna points to the typical line with a higher gain and transmits power to vehicles in other lines with a small gain. Therefore, the DL transmit power is the signal receiving power of  the vehicles, and the UL transmit power is the signal receiving power of the MBSs and SBSs. The DL transmit power can be written as
  \begin{equation}
     \vspace{-0.01cm}
     P_{r,V} = \left\{\begin{matrix}
       &{P_M}{G_{{M}}}{H_M}{\chi _M}{{\left\| x \right\|}^{ - {\alpha _M}}},&{x \in {\Phi _M}}\\
        & {P_S}{G_{S,0}}{H_{S,0}}{\chi _{S,0}}{\left\| x \right\|^{ - {\alpha _S}}},&x \in \Xi _{{l_0}}^{(S)}\\
     &{P_S}{G_{S,1}}{H_{S,1}}{\chi _{S,1}}{\left\| x \right\|^{ - {\alpha _{{S}}}}}, &	x \in {\Phi _S}\backslash \Xi _{{l_0}}^{(S)},
      \end{matrix}\right. \nonumber
   \end{equation}
   the UL  transmit power can be written as
   \begin{equation}
   \begin{split}
   {P_{r,M}}& = \begin{array}{*{20}{c}}
   {{P_V}{G_{V,1}}{H_{V,1}}{\chi _M}{{\left\| x \right\|}^{ - {\alpha _M}}},}&{x \in {\Phi _M}}
   \end{array},\\\nonumber
   {P_{r,S}} =& \left\{ {\begin{array}{*{20}{c}}
   	{{P_V}{G_{V,0}}{H_{V,0}}{\chi _{S,0}}{{\left\| x \right\|}^{ - {\alpha _S}}},}&{x \in \Xi _{{l_0}}^{(S)}}\\
   	{{P_V}{G_{V,1}}{H_{V,1}}{\chi _{S,1}}{{\left\| x \right\|}^{ - {\alpha _S}}},}&{x \in {\Phi _S}\backslash \Xi _{{l_0}}^{(S)}},
   	\end{array}} \right. \nonumber
  \end{split}
  \end{equation}
   where $ x $ is the distance between vehicle and BS. The $ P_{r,V} $, $  P_{r,M} $ and $ P_{r,S} $ are the signal receiving power of vehicle, MBS and SBS. $ {P_V} $, $ {P_M} $ and $ {P_S} $ are the transmit power of vehicle, MBS and SBS. $ {G_{{M}}} $, $ {G_{S,1}}$ and ${G_{S,0}} $ are the antenna gains of MBS, SBS that transmits power to vehicle in other lines, SBS that transmits power to vehicle in typical line. $ {G_{V,0}}$ and ${G_{V,1}} $ are the antenna gains of vehicle that transmits power to SBS and MBS. $ {\chi _M} $, $ {\chi _{S,0}} $ and $ {\chi _{S,1}} $ are the shadowing effects follow a log-normal distribution \cite{chetlur2019coverage}. The power-law path loss parameters are exponent $ {\alpha _M} $ and $ {\alpha _S} $ ($ \alpha > 2$). $ {H_M} $, $ {H_{S,0}} $, $ {H_{S,1}} $,  $ {H_{V,0}} $ and $ {H_{V,1}} $ are the channel fading gains of Nakagami-m with parameter $ m_{M} $, $ m_{S} $, $ m_{S,1} $, $ m_{V} $ and $ m_{V,1} $. The fading gains $H_i$, where $ i \in \{M, (S,0), (S,1), (V,0), (V,1) \} $, follow a Gamma distribution and its probability density function (PDF) is
    \begin{equation}
 f_{H_i}=\frac{m_{i}^{m_i}h^{m_i-1}}{\Gamma\left ( m_i \right ) }exp(-m_i h) \quad m \in \left \{ {m_M,m_{S,0},m_{S,1}} \right \}.\nonumber
    \end{equation}

\subsection{Association Policy for BSs and vehicles}

We assume that a vehicle is associated with the MBS/SBS which yields the highest biased received power in DL, similarly, the MBS/SBS will serve the vehicle with the highest biased transmit power in UL, namely the vehicles in the Voronoi cell of a BS connect to it. To handle the effect of shadowing, we use displacement theorem to express it as a random displacement of the location of the typical receiver \cite{chetlur2019coverage}. Thus, $ P_{r}(x)={P}{G}{\chi}{{\left\| x \right\|}^{ - {\alpha }}}  $ can be written as $ P_{r}(y)={P}{G_{{}}}{{\left\| y \right\|}^{ - {\alpha }}}  $, where $ y={\chi_{}^{-\frac{1}{\alpha }}}x $\textbf{}, then the transformed points form a two-dimensional (2D) homogeneous PPP with intensity $ \lambda $ to $ E\left [ {\chi_{}^{-\frac{2}{\alpha }}}  \right ]\lambda $, the one-dimensional (1D) PPP's intensity is $ \lambda $ to $ E\left [ {\chi_{}^{-\frac{1}{\alpha }}}  \right ]\lambda $ . Therefore, we use ${\lambda _L}$, ${\lambda _M}$, ${\lambda _S}$, ${\lambda _V}$ to denote the transformed ${\lambda _l}$, ${\lambda _m}$, ${\lambda _s}$, ${\lambda _v}$, respectively. In $ {\Phi _S}\backslash \Xi _{{l_0}}^{(S)} $, using the Theorem 1 in \cite{chetlur2019coverage}, the vehicles and SBSs form 2D PPP with intensity $ {\lambda _{Sa}} = E\left[ {\chi _{S,1}^{ - \frac{2}{\alpha }}} \right] {\pi} {\lambda _l}{\lambda _s} $, $ {\lambda _{Va}} = E\left[ {\chi _{S,1}^{ - \frac{2}{\alpha }}} \right] {\pi} {\lambda _l}{\lambda _v} $, respectively.

The typical vehicle associates with a MBS in DL if
 \begin{align}
  {P_M}{G_{{M}}}{B_M}{\left\| {{x_M}} \right\|^{ - {\alpha _M}}} > {P_S}{G_{S,0}}{B_{S}}{\left\| {{x_S}} \right\|^{ - {\alpha _S}}},\label{DL power}
 \end{align}
where $ B $ is selection bias factor. Otherwise, the vehicle connects to a SBS.

Similarly, the typical vehicle associates with a MBS in UL if
\begin{equation}
  P_V{G_{V,1}}{\left\| {{x_M}} \right\|^{ - {\alpha _M}}} > {P_V}{G_{V,0}}{\left\| {{x_S}} \right\|^{ - {\alpha _S}}}.\label{UL power}
\end{equation}
Otherwise, the typical vehicle associates with a SBS. We denote $ A_{M,S} = {P_M}{G_{{M}}{B_M}}/{P_S}{G_{S,0}}{B_S}$, $ {B_{M,S}} = {{P_V}{G_{V,1}}}/{P_V}{G_{V,0}}$, substituting for \eqref{DL power} and \eqref{UL power}:
\begin{align}
{A_{M,S}}{\left\| {{x_{{M}}}} \right\|^{ - {\alpha _M}}} > {\left\| {{x_{{S}}}} \right\|^{ - {\alpha _S}}},\label{assoc_A} \\
B_{M,S}{\left\| {{x_{{M}}}} \right\|^{ - {\alpha _M}}} > {\left\| {{x_{{S}}}} \right\|^{ - {\alpha _S}}},\label{assoc_B}
\end{align}
the transmit power of MBS is much larger than SBS and vehicle, $ A_{M,S} $ is larger than $B_{M,S}$.

\subsection{Interference for UL and DL}
In DL, the aggregate interference at the typical vehicle includes the interference from the MBSs $ I_M $, the interference from the SBSs in the typical line $ I_{S,0} $ and the interference from the SBSs in other lines $ I_{S,1} $. In UL, the aggregate interference includes the interference from the vehicles in the typical line $ I_{V,0} $ and the interference from the vehicles in the other lines $ I_{V,1} $. Therefore, the SINR in DL is
\begin{equation}
{SINR}_{D} = \frac{{{\mathop{ P}\nolimits} \left( {{{X}}^*} \right)}}{{{I_M} + {I_{S,0}} + {I_{S,1}} + {\sigma_D^2}}}.
\end{equation}
The SINR measured in UL is
\begin{equation}
{SINR}_{U} = \frac{{{\mathop{ P}\nolimits} \left( {{{X}}^*} \right)}}{{{I_{V,0}} + {I_{V,1}} + {\sigma_U^2}}},
\end{equation}
where $ {X}^* $ denotes the location with maximum received power.
Because the system is interference-limited and the noise is much smaller than interference \cite{chetlur2019coverage}, the thermal noise is neglected, $  {\sigma_U^2}= {\sigma_D^2}=0 $.

\section{SPECTRAL EFFICIENCY ANALYSIS}

In this section, we derive the joint association probabilities for UL/DL decoupled C-V2X using stochastic geometry. Based on the above analysis, the SE of each link in UL/DL decoupled access C-V2X can be derived.

\subsection{Association Probability}

According to the null probability of 1D and 2D PPP \cite{chetlur2019coverage}, the cumulative distribution functions (CDF) of $ x_S $, $ x_M $ are
\begin{align}
{F_S}\left( {{x_S}} \right) &= 1 - \exp \left( { - 2{\lambda _S} {x_{{S}}}} \right),\\
{F_M}\left( {{x_M}} \right) &= 1 - \exp \left( { - {\lambda _M}\pi x_M^2} \right).\label{cdf_2}
\end{align}
According to $ f(x)={\mathrm{d}F(x) }/{\mathrm{d} x} $, the probability density functions (PDF) of $ x_S $, $ x_M $ are
\begin{align}
\label{pdf_1}{f_S}\left( {{x_S}} \right) &= 2{\lambda _S}\exp \left( { -2 {\lambda _S} {x_{{S}}}} \right),\\
\label{pdf_2}{f_M}\left( {{x_M}} \right) &= 2\pi {x_M}{\lambda _M}\exp \left( { - {\lambda _M}\pi x_M^2} \right).
\end{align}
According to the association policy given by \eqref{assoc_A} and \eqref{assoc_B}, we derive the joint association probability as follows:

1) Case 1: The probability that the typical vehicle connects to MBSs both in DL and UL is
\begin{align}
 \Pr \left( {{{Case }} \,1} \right)=&\Pr \left( {{{{A}}_{M,S}}X_M^{^{ - {\alpha _M}}} > X_S^{^{ - {\alpha _S}}};} \right. \nonumber\\
&\quad \left. {{B_{M,S}}X_M^{^{ - {\alpha _M}}} > {X_S}^{ - {\alpha _S}}} \right).
\end{align}
 According to the consideration of ${{{A}}_{M,S}} > {B_{M,S}}$, the above formula can be written as
 \begin{align}
 \Pr \left( {{{Case }} \, 1} \right) = \Pr \left( {{X_{{M}}} < B_{M,S}^{\frac{1}{{{\alpha _M}}}}X_S^{\frac{{{\alpha _S}}}{{{\alpha _M}}}}} \right).
 \end{align}
The joint association probability of Case 1 is
\begin{align} \label{pr_1}
&\Pr \left( {{{Case}} \,1} \right) \nonumber \\
&=1 - \int_0^\infty  {\left[ {2{\lambda _S}\exp \left( { - {\lambda _M}\pi B_{M,S}^{\frac{2}{{{\alpha _M}}}}x_S^{\frac{{2{\alpha _S}}}{{{\alpha _M}}}}- 2{\lambda _S}{x_{{S}}}} \right)} \right]} d{x_{{S}}},
\end{align}
when ${\alpha _S} = {\alpha _M}$, $ \Pr \left(Case\,1\right) $ in closed form is
\begin{equation}\label{cloesed form}
\begin{split}
&\Pr \left( {{{Case }} \,1} \right)=\\
& 1 - \sqrt {\frac{{\lambda _S^2}}{{{\lambda _M}B_{M,S}^{\frac{2}{\alpha }}}}} \exp \left( {\frac{{\lambda _S^2}}{{{\lambda _M}\pi B_{M,S}^{\frac{2}{\alpha }}}}} \right)erfc\left( {\frac{{{\lambda _S}}}{{\sqrt {{\lambda _M}\pi B_{M,S}^{\frac{2}{\alpha }}} }}} \right).
\end{split}
\end{equation}

2) Case 2: The probability that the typical vehicle connects to SBS in UL and MBS in DL is
\begin{align}
&\Pr \left({{{Case \,2}}} \right)\nonumber\\
&= \Pr \left( {{{{A}}_{M,S}}X_M^{^{ - {\alpha _M}}} > X_S^{^{ - {\alpha _S}}};{B_{M,S}}X_M^{^{ - {\alpha _M}}} < X_S^{^{ - {\alpha _S}}}} \right)\nonumber\\
&= \Pr \left( {{B_{M,S}}X_M^{^{ - {\alpha _M}}} < X_S^{^{ - {\alpha _S}}} < {{{A}}_{M,S}}X_M^{^{ - {\alpha _M}}}} \right).
\end{align}
The joint association probability of Case 2 is formulated as
\begin{equation}
\begin{aligned}
&\Pr \left( {Case} \,2 \right)\\
&= \int_0^\infty  {\left[ {2{\lambda _S}\exp \left( { - {\lambda _M}\pi B_{M,S}^{\frac{2}{{{\alpha _M}}}}x_S^{\frac{{2{\alpha _S}}}{{{\alpha _M}}}} - 2{\lambda _S}{x_{{S}}}} \right)} \right]} d{x_{{S}}}\\
&\quad - \int_0^\infty  {\left[ {2{\lambda _S}\exp \left( { - {\lambda _M}\pi A_{M,S}^{\frac{2}{{{\alpha _M}}}}x_S^{\frac{{2{\alpha _S}}}{{{\alpha _M}}}} - 2{\lambda _S}{x_{{S}}}} \right)} \right]} d{x_{{S}}}.
\end{aligned}
\end{equation}

3) Case 3: The probability that the typical vehicle connects to MBS in UL and SBS in DL is
\begin{align}\label{case3}
\Pr \left( {{{Case }} \, 3} \right)=&\Pr \left( {{{{A}}_{M,S}}X_M^{^{ - {\alpha _M}}} < X_S^{^{ - {\alpha _S}}};} \right.\nonumber\\
&\quad \left. {{B_{M,S}}X_M^{^{ - {\alpha _M}}} > {X_S}^{ - {\alpha _S}}} \right),
\end{align}
because ${{{A}}_{M,S}} > {B_{M,S}}$, the domain in $ \eqref{case3} $ is empty, $ \Pr \left( {{{Case }} \, 3} \right)= 0  $, which means that Case 3 won't happen. Therefore, we will not discuss Case 3 in later sections.

4) Case 4: The probability that the typical vehicle connects to SBS both in UL and DL is:
\begin{align}\label{case4}
\Pr \left( {{{Case }} \, 4} \right)=&\Pr \left( {{{{A}}_{M,S}}X_M^{^{ - {\alpha _M}}} < X_S^{^{ - {\alpha _S}}};} \right.\nonumber\\
&\quad \left. {{B_{M,S}}X_M^{^{ - {\alpha _M}}} < {X_S}^{ - {\alpha _S}}} \right)\nonumber\\
& = \Pr \left( {{X_{{S}}} < A_{S,M}^{\frac{1}{{{\alpha _S}}}}X_M^{\frac{{{\alpha _M}}}{{{\alpha _S}}}}} \right).
\end{align}
The joint association probability of Case 4 is formulated as
\begin{align}
&\Pr \left( {{{Case \, 4}}} \right) \nonumber\\
&= \int_0^\infty  {\left[ {2{\lambda _S}\exp \left( { - {\lambda _M}\pi A_{M,S}^{\frac{2}{{{\alpha _M}}}}x_S^{\frac{{2{\alpha _S}}}{{{\alpha _M}}}} - 2{\lambda _S}{x_{{S}}}} \right)} \right]} d{x_{{S}}}.
\end{align}
\begin{IEEEproof}
	Due to space constraints, we take the proof of Case 1 as an example.
	The proof of joint association probability of Case 1 (UL=DL=MBS) is
	\begin{align}
		& \Pr \left( {{{Case }} \,1} \right) =Pr ({B_{M,S}}X_M^{^{ - {\alpha _M}}} > X_S^{^{ - {\alpha _S}}})\nonumber\\
		&= {{{{{E}}_{{X_S}}}}}\left[ {\Pr \left( {{X_{{M}}} < B_{M,S}^{\frac{1}{{{\alpha _M}}}}x_S^{\frac{{{\alpha _S}}}{{{\alpha _M}}}}\left| {{X_S}} \right.} \right)} \right]\nonumber\\
		&=\int_0^\infty  {{F_M}\left( {B_{M,S}^{\frac{1}{{{\alpha _M}}}}x_S^{\frac{{{\alpha _S}}}{{{\alpha _M}}}}} \right)} {f_S}\left( {{x_{{S}}}} \right)d{x_{{S}}}\nonumber\\
		& \overset{(a)}{=}1 - \int_0^\infty  {\left[ {2{\lambda _S}\exp \left( { - {\lambda _M}\pi B_{M,S}^{\frac{2}{{{\alpha _M}}}}x_S^{\frac{{2{\alpha _S}}}{{{\alpha _M}}}} - 2{\lambda _S}{x_{{S}}}} \right)} \right]} d{x_{{S}}},
	\end{align}

	where (a) follows from the substituting $ F_{M}\left ( \cdot  \right ) $ and $ f_{S}\left ( \cdot  \right ) $ from Eq. \eqref{cdf_2} and Eq. \eqref{pdf_1} in the previous steps. And the special case ${\alpha _S} = {\alpha _M}$ can follow the proof of Lemma 2 in \cite{chetlur2019coverage}.
	
	Similarly, the proofs of other joint association probabilities can be derived by following the steps as Case 1.	
\end{IEEEproof}

\subsection{Distance Distribution of A Vehicle to Serving BSs}

The exact expressions of distance distributions to serving BSs of a vehicle for three cases are derived in this section. Because Case 1's and Case 4's UL and DL are associated with the same type BSs, Case 1's and Case 4's distance distributions are only one type. While Case 2's UL and DL connect to different types of BSs, there are two types of distance distributions to consider.

1) The distance distribution of Case 1 is formulated as
\begin{align}\label{pdfM1}
{f_{{X_M}\left| {{{Case }}\,1} \right.}} = \frac{{\exp \left( { - {\lambda _S}\pi 2B_{S,M}^{\frac{1}{{{\alpha _S}}}}{x^{\frac{{{\alpha _M}}}{{{\alpha _S}}}}}} \right){f_{{X_M}}}\left( x \right)}}{{\Pr \left( {{{Case }}\,1} \right)}}.
\end{align}

2) The distance distribution of Case 2 is formulated as
\begin{align}
{f_{{X_M}\left| {{{Case}}~ {{2}}} \right.}} =& \frac{{{f_{{X_M}}}\left( x \right)}}{{\Pr \left( {{{Case}}~ {{2}}} \right)}}\left[ {\exp \left( { - {\lambda _S}\pi 2A_{S,M}^{\frac{1}{{{\alpha _S}}}}{x^{\frac{{{\alpha _M}}}{{{\alpha _S}}}}}} \right)} \right. - \nonumber\\
&\left. {\exp \left( { - {\lambda _S}\pi 2B_{S,M}^{\frac{1}{{{\alpha _S}}}}{x^{\frac{{{\alpha _M}}}{{{\alpha _S}}}}}} \right)} \right],
\label{dis xm case2}
\end{align}

\begin{align}
{f_{{X_S}\left| {{{Case }}~2} \right.}} = &\frac{{{f_{{X_S}}}\left( x \right)}}{{\Pr \left( {{{Case~ 2}}} \right)}}\left[ {\exp \left( { - {\lambda _M}\pi B_{M,S}^{\frac{2}{{{\alpha _M}}}}{x^{\frac{{{\alpha _S}}}{{{\alpha _M}}}}}} \right) - } \right.\nonumber\\
&\left. {\exp \left( { - {\lambda _M}\pi A_{M,S}^{\frac{2}{{{\alpha _M}}}}{x^{\frac{{{\alpha _S}}}{{{\alpha _M}}}}}} \right)} \right].
\end{align}

3) The distance distribution of Case 4 is formulated as
\begin{align}\label{pdf4}
{f_{{X_{{S}}}\left| {{{Case }} \,4} \right.}} = \frac{{\exp \left( { - {\lambda _{{M}}}\pi {{A}}_{M,S}^{\frac{2}{{{\alpha _M}}}}{x^{\frac{{2{\alpha _S}}}{{{\alpha _M}}}}}} \right){f_{{X_S}}}\left( x \right)}}{{\Pr \left( {{{Case}}\, 4} \right)}}.
\end{align}

\begin{IEEEproof}
	The detailed proofs of the distance distribution are similar to Lemma 4 in \cite{sattar2019spectral}. The proof is omitted due to space constraints.
\end{IEEEproof}

\subsection{Spectral Efficiency}
\textit{Theorem 1:} Based on the results of \eqref{pdf_1}, \eqref{pdf_2} and \eqref{pr_1} to \eqref{pdf4}, we take the SE of Case 2 UL as an example and the expression can be given by
\begin{align}\label{se2u}
\tau _{2}^U = \int_0^\infty  {{f_{S\left| {{2}} \right.}}\int_0^\infty  {\Pr \left[ {{H_{S,0}} > {\beta _{S,0}}{I_{U,2}}x_S^{{\alpha _S}}} \right]} dtd{x_S}},
\end{align}
where $ {\beta _{S,0}} = \left({e^t} - 1\right)  / \left({P_V}{G_{S,0}}\right) $, $ {I_{U,2}} = {I_{S,0}} + {I_{S,1}} $. For the convenience of writing, we use $ {f_{S\left| {{{2   }}} \right.}}\left ( \cdot  \right ) $  to substitute $ {f_{{X_S}\left| {{{Case\,2}}} \right.}}\left ( \cdot  \right )$ and the following PDFs are simplified to the same form.
\begin{IEEEproof}
	The SE of Case 2 UL is
	\begin{align}
		&\tau _{2}^U=E\left[ {ln \left( {1 + {{SIN}}{{{R}}_{U,S}}} \right)} \right]\nonumber\\
		&= \int_0^\infty  {{f_{S\left| {{2}} \right.}}\int_0^\infty  {\Pr \left[ {ln \left( {1 + \frac{          P\left( {{{X}}^*} \right) }{{{I_{U,2}}}}} \right) > t} \right]} dtd{x_S}} \nonumber\\
		&\mathop {{{  =  }}}\limits^{} \int_0^\infty  {{f_{S\left| {{2}} \right.}}\int_0^\infty  {\Pr \left[ {{H_{V,0}} > \frac{{\left( {{e^t} - 1} \right){I_{U,2}}}}{{{P_V}{G_{V,0}}}}x_S^{{\alpha _S}}} \right]} dtd{x_S}}\nonumber \\
		&\mathop  = \limits^{(a)} \int_0^\infty  {{f_{S\left| {{2}} \right.}}\int_0^\infty  {\Pr \left[ {{H_{V,0}} > {\beta _{S,0}}{I_{U,2}}x_S^{{\alpha _S}}} \right]} dtd{x_S}},
	\end{align}
	where $P\left( {{X}}^*\right)={{P_V}{G_{V,0}}{H_{V,0}}x_S^{{-\alpha _S}}} $, $ {\beta _{S,0}} = \left({e^t} - 1\right)  / \left({P_V}{G_{V,0}}\right) $  in  (a).
	The proof of $  \Pr \left[ {{H_{V,0}} > {\beta _{S,0}}{I_{U,2}}x_S^{{\alpha _S}}} \right] $ is
	\begin{align}
		&\Pr \left[ {{H_{V,0}} > {\beta _{S,0}}{I_{U,2}}x_S^{{\alpha _S}}} \right]\nonumber\\
		&= {E_{{I_S}}}\left\{ {\Pr \left[ {{H_{V,0}} > {\beta _{S,0}}{I_{U,2}}x_S^{{\alpha _S}}} \right]} \right\}\nonumber\\
		&\mathop  = \limits^{({{a}})} {E_{{I_S}}}\left[ {\frac{{\Gamma \left( {{m_S},{m_S}{\beta _{S,0}}{I_{U,2}}x_S^{{\alpha _S}}} \right)}}{{\Gamma \left( {{m_S}} \right)}}} \right]\nonumber\\
		&\mathop  = \limits^{(b)} {E_{{I_S}}}\left[ {\sum\limits_{k = 0}^{{m_S} - 1} {\frac{{{{\left( {{m_S}{\beta _{S,0}}{I_{U,2}}x_S^{{\alpha _S}}} \right)}^k}}}{{k!}}{e^{ - {m_S}{\beta _{S,0}}{I_S}x_S^{{\alpha _S}}}}} } \right]\nonumber\\
		&= \sum\limits_{k = 0}^{{m_S} - 1} {\frac{{{{\left( { - {m_S}{\beta _{S,0}}x_S^{{\alpha _S}}} \right)}^k}}}{{k!}}{{\left[ {\frac{{{\delta ^k}}}{{\delta {j^k}}}{\zeta _{_{{I_{U,2}}}}}\left( j \right)} \right]}_{j = {m_S}{\beta _{S,0}}x_S^{{\alpha _S}}}}},
	\end{align}
	where (a) follows from the CCDF of gamma random variable
	$ H_{V,0} $, and (b) follows from the definition of incomplete
	gamma function for integer values of $ m_S $. The aggregate interference can be divided into two independent components $ I_{V,0}, I_{V,1} $, the Laplace transform of interference can be computed as product of the Laplace transforms of the two components, thus, $ {\zeta _{{I_{U,2}}}}\left( j \right) = {\zeta _{_{{I_{V,0}}}}}\left( j \right){\zeta _{_{{I_{V,1}}}}}\left( j \right) $. Similar to the proof of Lemma 5 in \cite{chetlur2019coverage}, the Laplace transforms of $ I_{V,0} $ and $ I_{V,1} $ are
	\begin{align}
		&{\zeta _{{I_{V,0}}}}\left( j \right)= {E_{{I_{V,0}}}}\left[ {\exp \left( { - j{I_{V,0}}} \right)} \right]\nonumber\\
		&= {E_{{I_{V,0}}}}\left[ {\exp \left( { - j\sum\limits_{{x} \in \Xi _{{l_0}}^{V}\backslash [ - {x_S},{x_S}]} {{P_V}{G_{V,0}}{H_{V,0}}{x^{ - {\alpha _S}}}} } \right)} \right]\nonumber\\
		&\mathop  = \limits^{(a)} {E_{\Xi _{{l_0}}^{V}\backslash {{X}}^*}}{E_{{H_{V,0}}}}\left[ {\prod\limits_{{x} \in \Xi _{{l_0}}^{V}\backslash [ - {x_S},{x_S}]} {{e^{ - j{P_V}{G_{V,0}}{H_{V,0}}{x^{ - {\alpha _S}}}}}} } \right]\nonumber\\
		&\mathop  = \limits^{(b)} \exp \left[ { - 2{\lambda _V}\int_{{x_S}}^\infty  {\left(1 - {{\left( {1{{ + }}\frac{{j{P_V}{G_{V,0}}{x^{ - {\alpha _S}}}}}{{{m_S}}}} \right)}^{ - {m_S}}}\right)dx} } \right],
	\end{align}
	where (a) follows from the independence of  $ \Xi _{{l_0}}^{V} $, we convert the accumulative form into accumulative multiplication. (b) follows from the Nakagami-m fading assumption and the PGFL of a 1D PPP. Similarly, the proof of $ \zeta _{{I_{S,1}}} $ can be derived by following the same steps.
\end{IEEEproof}	
\begin{figure}[t]		
	\centering
	\subfigure[LOS]{\label{pr-los}\includegraphics[width=0.84\hsize]{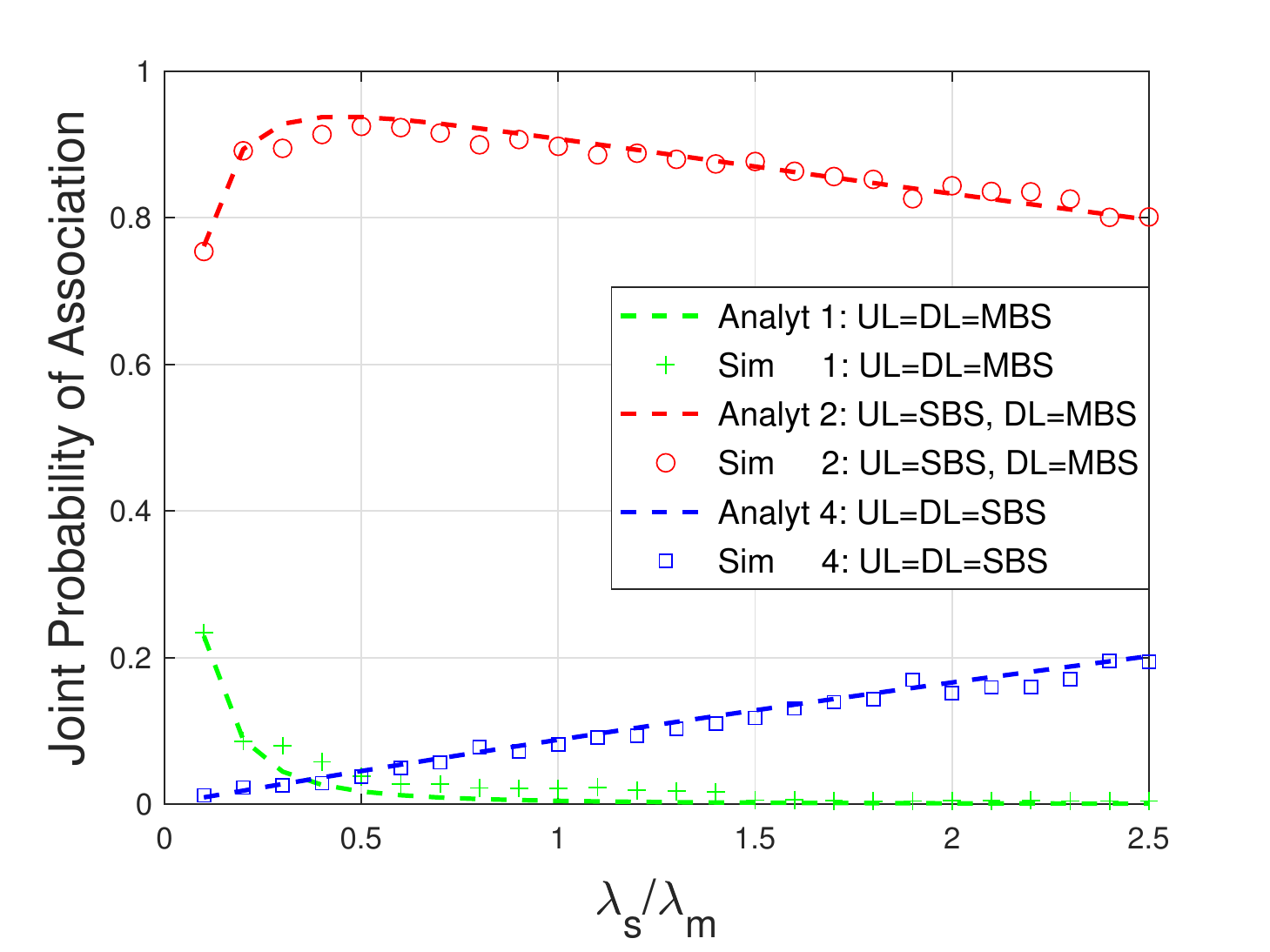}}
	\subfigure[NLOS]{\label{pr-nlos}\includegraphics[width=0.84\hsize]{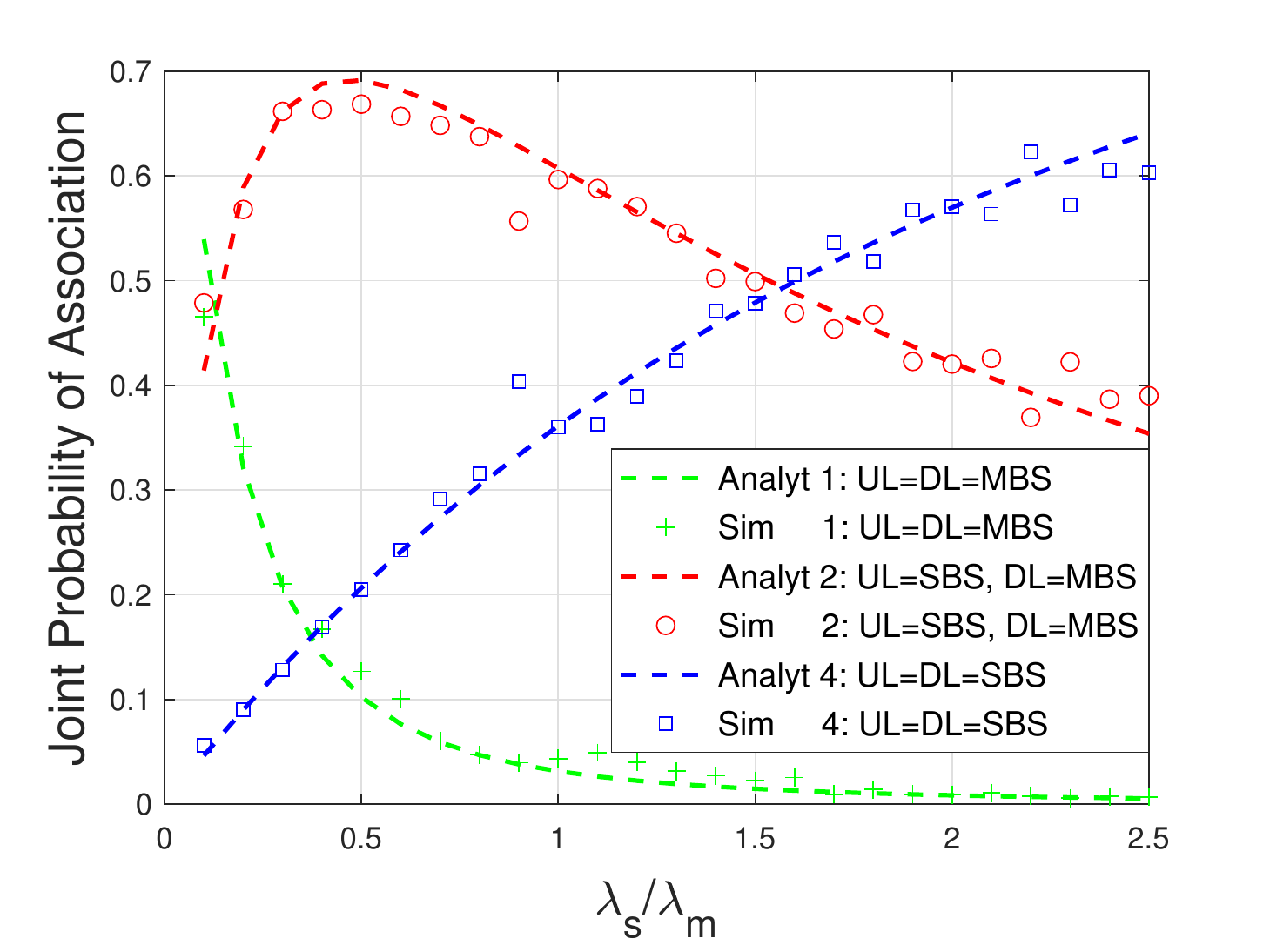}}	
	\caption{Joint association probability for Cases $ 1-4 $ ( $P_{M} = 46~ dBm; P_{S} = 20 ~dBm; P_{V} = 20~dBm$ ).}
	\label{pr}
\end{figure}

\begin{figure}[t]	
	\centering
	\subfigure[LOS]{\label{se-los} \includegraphics[width=0.84\hsize]{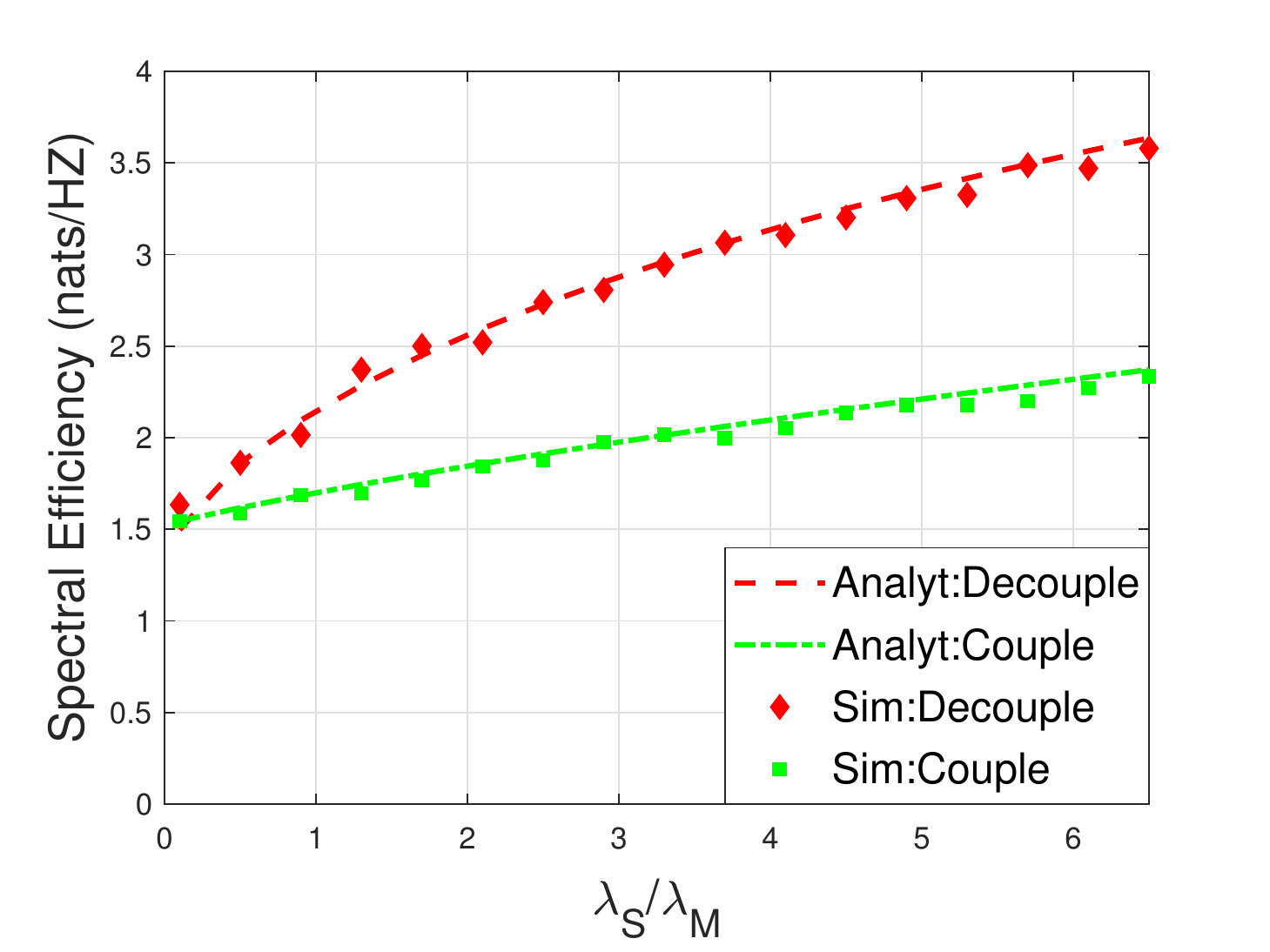}}
	\subfigure[NLOS]{\label{se-nlos} \includegraphics[width=0.84\hsize]{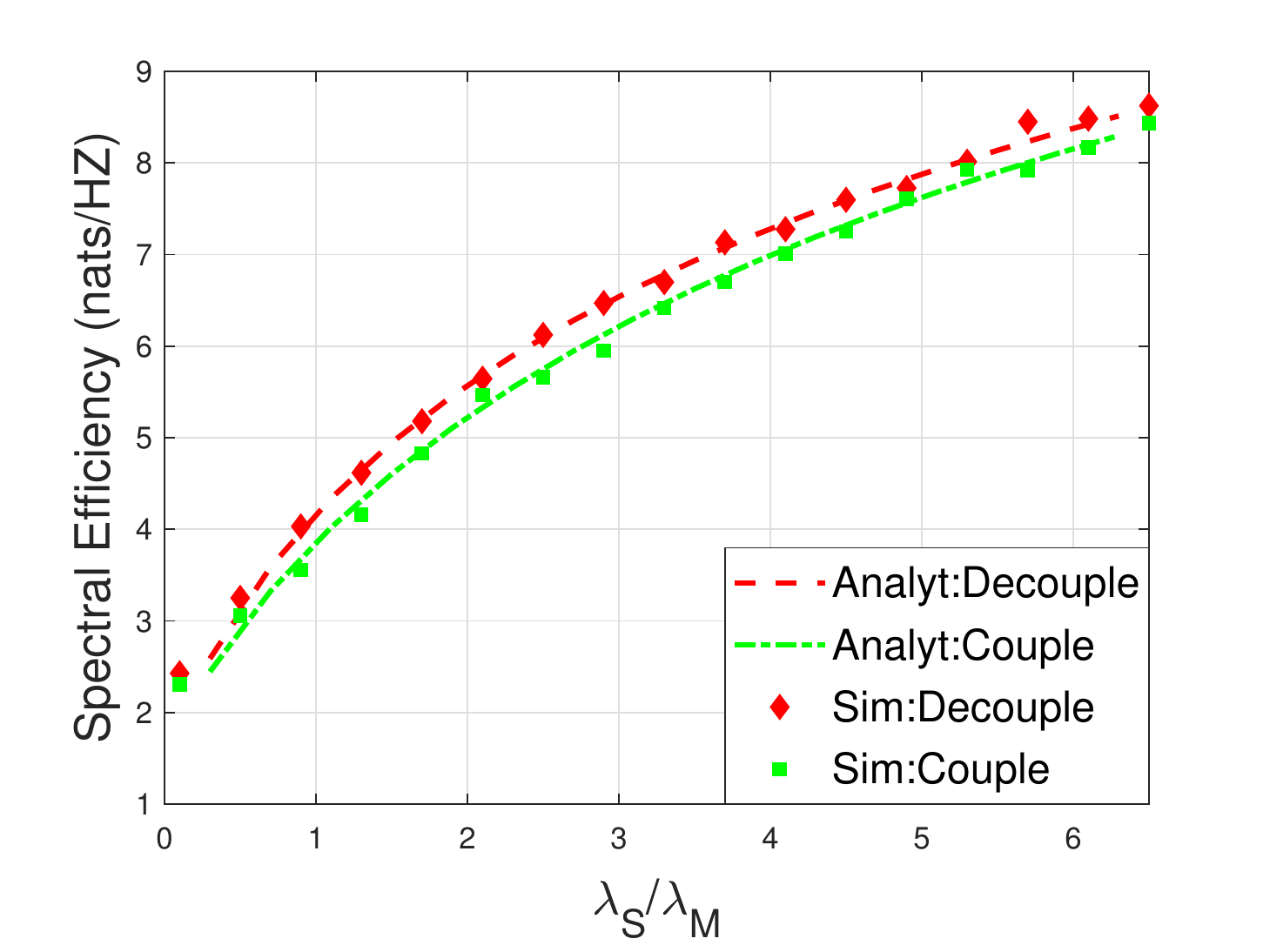}}		
	\caption{System average SE for decoupled and coupled cases in LOS and NLOS scenarios.}
	\label{se}
\end{figure}

Based on the results of \eqref{pr_1} to \eqref{se2u}, the other links' expressions of SE can be derived by following the similar steps of Theorem 1 in UL/DL decoupled C-V2X.

The average system SE of UL/DL decoupled C-V2X is
\begin{equation}
SE = \sum_{i=1}^{4} \sum_{n}^{\left \{ U,D \right \}}\tau _{{Case\, i}}^{{n}}{Pr}\left ( Case~ i \right ),
\end{equation}
where $ \tau _{{Case\, i}} $ is the SE of Case i's n-link, here $ i \in \left\{ {1, 2, 4} \right\} $, $ n \in \left\{ {U, D} \right\} $.

The SE of coupled access could be derived as Corollary 2 in \cite{sattar2019spectral}.

\section{NUMERICAL RESULTS}

In this section, more than 100,000 Monte Carol simulations are executed in a UL/DL decoupled access C-V2X scenario that the simulation observation area is a circular area with a radius of 5 km for line of sight (LOS) and non line of sight (NLOS). 
Table \ref{sys-par} shows the default system parameters \cite{sattar2019spectral}. We use sim and analyt to denote simulated and analytical in legend to save the space in figures.

\begin{table}[h]
	\centering
	\caption{SYSTEM PARAMETERS}	\label{sys-par}
	\begin{tabular}{l|l }
		\toprule
		\hline
		\label{table1}
		Parameters  & Value \\
		\hline
		Macro BS transmit power $P_{M}$ (dBm)& 46 \\
		Small BS transmit power $P_{S}$ (dBm)& 20 \\
		Vehicle transmit power $P_{V}$ (dBm)& 20 \\
		Pathloss exponent for MBS  $\alpha _{M}$ &4 \\
		Pathloss exponent for SBS $\alpha _{S}$(NLOS) &4 \\
		Pathloss exponent for SBS $ \alpha _{S}$(LOS) &2 \\
		Antenna Gain for MBS $G_{M}$(dBi) &0 \\
		Antenna Gain for vehicle in typical line $G_{S0}$ (dBi) &0 \\
		Antenna Gain for vehicle in other line $G_{S1}$ (dBi) &-20 \\
		Line density  $ \lambda_{l}$  (1/km)   &10  \\
		\hline
		\bottomrule
	\end{tabular}
\end{table}

We first analyze the joint association probabilities of three cases as shown in Fig. \ref{pr-los} and Fig. \ref{pr-nlos}.
The joint association probability is equal to the percentage of vehicles that choose the particular case.
We observe that in UL/DL decoupled C-V2X networks, the probability of vehicles choosing Case 2 to communicate increases rapidly at the beginning, then decreases slowly at the cost of the increase in association Case 4 i.e., when vehicles choose to connect with SBSs in both UL and DL, as the ratio of the density of SBS and MBS increases in both LOS and NLOS.
The probability of Case 4 is increasing all the time in both LOS and NLOS. The reason behind this phenomenon is obvious, with the increasing density of SBS,
the SBSs are getting closer to the vehicles and the closer distance makes up for the disadvantage of lower power compared with MBS.

Fig. \ref{se} shows the main result i.e., the comparison of the decoupled and coupled average system SE in LOS and NLOS. It shows that the SE is improved both in LOS and NLOS after being decoupled. Specifically, we can see that the improvement is increasing as the ratio of the density of SBS and MBS increases in LOS and the performance is nearly increased by 50\% in LOS. While in NLOS, the improvement is very small.
UL/DL decoupled access separates UL from DL, UL could be connected to the BS with the largest signal power, but not the DL'BS that may not have the largest signal power in UL. Therefore, the improvement of SE mainly comes from UL.
In NLOS, according to Fig. \ref{pr-los}, most vehicles choose Case 2 i.e., when vehicles choose to connect with MBS in
DL and SBS in UL, thus the SE improvement comes from UL'SBS and most vehicles connect to MBS in coupled mode, thus SE has a higher improvement as the ratio of SBS and MBS increase. While in NLOS, most vehicles choose Case 4 as the SBS increases, therefore, the improvement which comes from Case 2's UL is very small, thus the system average SE's improvement is not obvious. This phenomenon suggests that UL/DL decoupled access could bring gains compared to traditional coupled access and this technology can be used as a tool for achieving higher SE, especially in the LOS scenario in C-V2X.

\vspace{-0.01cm}

\section{Conclusion}

In this paper, we have proposed a UL/DL decoupled C-V2X networks analytical framework and have analyzed the SE of the UL/DL between vehicles and MBSs/SBSs. Specifically, we have analyzed the joint association probability, distance distributions to its serving BSs and the SE of all links in UL/DL decoupled C-V2X networks. Theoretical analysis and empirical results have verified the effectiveness of the proposed analytical framework and demonstrated that with UL/DL decoupled access in C-V2X, better system average SE can be achieved than coupled access both in LOS and NLOS. For the future work, we will focus on the mobility management and joint access optimization in UL/DL decoupled C-V2X.

\vspace{0.01cm}
\section*{Acknowledgment}

This work is supported in part by the National Natural Science
Foundation of China (No. 61871211), in part by the Summit of the Six Top Talents Program of Jiangsu Province and in part by the Natural Science Fund for Distinguished Young Scholars of Jiangsu Province under Grant BK20220067.

\bibliographystyle{IEEEtran}

\bibliography{references}

\end{document}